\documentclass[twocolumn]{aastex631}

\usepackage[none]{hyphenat} 
\usepackage{comment}
\usepackage{eqnarray,amsmath}

\newcommand{\lya}{Ly$\alpha$}

\newcommand{\grizli}{{\tt\string GRIZLI}}

\newcommand{\muv}{\(\rm M_{UV}\)}
\newcommand{\ha}{H$\alpha$}
\newcommand{\oiii}{[O\thinspace{\sc iii}]}
\newcommand{\galid}{772319}

\begin{document}

\title{Joint Survey Processing. III. Compact Oddballs in the COSMOS Field - Little Red Dots and Transients}

\author[0000-0001-8792-3091]{Yu-Heng Lin}
\affiliation{Caltech/IPAC, 1200 E. California Blvd. Pasadena, CA 91125, USA}
\email{ianlin@ipac.caltech.edu}

\author[0000-0002-9382-9832]{Andreas L. Faisst}
\affiliation{Caltech/IPAC, 1200 E. California Blvd. Pasadena, CA 91125, USA}

\author[0000-0001-7583-0621]{Ranga-Ram Chary}
\affiliation{SPACE Institute, University of California, 603 Charles E. Young Drive East,Los Angeles, CA 90095-1567}

\author[0000-0002-6610-2048]{Anton M. Koekemoer}
\affiliation{Space Telescope Science Institute, 3700 San Martin Drive, Baltimore, MD 21218, USA}

\author[0000-0003-2638-720X]{Joseph Masiero}
\affiliation{Caltech/IPAC, 1200 E. California Blvd. Pasadena, CA 91125, USA}

\author[0000-0001-5382-6138]{Daniel Masters}
\affiliation{Caltech/IPAC, 1200 E. California Blvd. Pasadena, CA 91125, USA}

\author[0000-0001-7166-6035]{Vihang Mehta}
%\affiliation{IPAC, Mail Code 314-6, California Institute of Technology, 1200 E. California Blvd., Pasadena, CA, 91125, USA}
\affiliation{Caltech/IPAC, 1200 E. California Blvd. Pasadena, CA 91125, USA}

\author[0000-0002-7064-5424]{Harry I. Teplitz}
\affiliation{Caltech/IPAC, 1200 E. California Blvd. Pasadena, CA 91125, USA}

\author[0000-0002-6313-6808]{Gregory L. Walth}
\affiliation{Caltech/IPAC, 1200 E. California Blvd. Pasadena, CA 91125, USA}

\author[0000-0003-1614-196X]{John R. Weaver}
\affiliation{Department of Astronomy, University of Massachusetts, Amherst, MA 01003, USA}

\begin{abstract}
We present the HST ACS G800L grism spectroscopy observation of the faint active galactic nuclei (AGN) candidates in the COSMOS field at redshift of 6 selected by the point-source morphology and the photometry drop-off at 8000\AA.  Among the sample of 7 objects, only one is detected by multiple bands, and has similar shape of spectral energy distribution as the so-called ``little red dots'' JWST selected AGN candidates, but our object is 3 magnitude brighter than the JWST sample.  We draw the upper limit of the AGN luminosity function $\Phi=1.1\times 10^{-7}$Mpc$^3$ mag$^{-1}$ for \muv=$-21$ at redshift of 6.  
The rest of the sample shows inconsistent flux density when comparing magnitudes of HST ACS F814W to the Subaru $i$-band and $z$-band magnitudes combined. The HST ACS G800L grism observation shows that this inconsistency cannot be created from an emission line. Therefore, we speculate that these objects are transients with the light curve decay timescale at most 6 years in observed frame. 
%RC: Most likely asteroids given the -10 deg ecliptic latitude of COSMOS. 

\end{abstract}

\keywords{galaxies: peculiar, galaxies: high-redshift, galaxies: photometry,}

%%%%%%%%%%%%%%%%%%%%%%%%%%%%%%%%%%%%%%%%%%%%%%%%%%%%%
%%%%%%%%%%%%%%%%%%%%%%%%%%%%%%%%%%%%%%%%%%%%%%%%%%%%%
\section{Introduction} \label{sec:intro}
%%%%%%%%%%%%%%%%%%%%%%%%%%%%%%%%%%%%%%%%%%%%%%%%%%%%%
%%%%%%%%%%%%%%%%%%%%%%%%%%%%%%%%%%%%%%%%%%%%%%%%%%%%%

The existence of Quasi-stellar objects (QSOs) or quasars at high redshift in the Epoch of Reionization (EoR, $z>6$) raises the question of how the central massive black hole builds up its mass in such a short time, how the active galactic nucleus (AGN) impacts the growth of the host galaxy, and how common does this mode happen among all the high redshift galaxies.  
Tracing the evolution of the quasar UV luminosity function (LF), and especially the AGNs at the faint end (\muv$>$ -23), provides potential insights to understand the origin of these systems. 
Prior to the lunch of the James Webb Space Telescope (JWST), the quasar LF at $z<4$ has been measured and often expressed by the double power law formula \citep{Boyle_1988, Pei_1995,Croom_2004, Richards_2006, Masters_2012}.  At higher redshift, the bright end of the quasar LF (\muv$< -22$) has been measured \citep{McGreer_2018, Matsuoka_2018, Kulkarni_2019,Niida_2020}, while the faint end of the LF at $z>5$ is less understood \citep{Kulkarni_2019, Giallongo_2019,Morishita_2020,Faisst_2022}.

Typically, identifying faint AGNs mainly relies on their point-source morphology and the distinct rest-frame UV radiation features, including the sudden drop across the Lyman or \lya\ break \citep{Akiyama_2018, Niida_2020}.
The challenge in these techniques is it requires at least a magnitude deeper in the blue band to measure the Lyman/\lya\ break, and high spatial resolution to distinguish galaxies and point sources.  The point sources may also be contaminated by faint cool dwarfs if their photometry peaks at the wavelength same as the targeted observed Lyman break \citep{Fajardo-Acosta_2016}, or confused by high redshift transients if the blue and red bands are observed at different times \citep{Yasuda_2019}.

With the aid of the remarkable sensitivity in the mid-infrared, JWST is able to select quasar and AGN candidates through their rest-frame optical emission \citep{Harikane_2023, Kocevski_2023,Labbe_2023, Oesch_2023, Onoue_2023}. A particular intriguing discovery from JWST is the compact objects, called ``Little Red Dots'' (LRDs), with a unique ``v-shape'' spectral energy distribution (SED) -- a blue UV continuum and a steep red optical continuum \citep{Labbe_2023,Matthee_2024, Greene_2024, Kokorev_2024, Akins_2024}, where the red colors in the optical bands are consistent with either a reddened AGN or dusty star formation, with substantial dust attenuation \citep[A$_{V}\sim 1-3$, ][]{Kokorev_2024, Greene_2024}.   
A small sample of these LRDs have been followed up with spectroscopy \citep{Matthee_2024, Greene_2024}, and 80$\%$ of the objects present broad \ha\ emission lines and the absent of broad \oiii\ emission that can only be explained by AGN-dominated SED \citep{Greene_2024, Maiolino_2024}. 
The conflict of the ``v-shape'' SED is that the red optical continuum indicate a moderate levels of dust attenuation, which is inconsistent with the blue UV continuum. 
%This "v-shape" SED has no lower redshift counterpart {\color{red} [cite]}. 
On top of the puzzling SED, the density of the LRDs are found 10$\sim$100 times more abundant than the UV-selected faint AGN \citep{Labbe_2023,Matthee_2024,Greene_2024,Kokorev_2024,Akins_2024}.

Spectroscopic follow-up of these faint AGN candidates will provide us more information to reveal their ambiguous nature. 
For example, \lya~emission is commonly used to study a galaxy's ability to reionize the surrounding intergalactic medium (IGM), but its flux in faint AGNs remains poorly constrained.
%The \lya~emission, in particular, is a proxy for the faint AGNs' ability to reionized the surrounding intergalactic medium (IGM).
The challenge of revealing these questions comes from the their faint luminosities, with the average observed magnitude $\sim$27.70 mag at 1$\mu$m and $\sim$27.40 mag at 2$\mu$m \citep{Kokorev_2024}.  Therefore, the brightest object among the LRDs become the best candidate to follow up.

\citet{Faisst_2022} selected $12$ faint AGN candidates over the total field of the Cosmic Evolution Survey \citep[COSMOS;][]{Koekemoer_2007, Scoville_2007}.  These objects were selected by their compact size ($r_e < 0.08$\arcsec) and the color difference between HST ACS F814W \citep{Scoville_2007} and Subaru Hyper-SuprimeCam \citep[HSC, ][]{Miyazaki_2018} $i$-band \citep{Aihara_2018}.
The limit in size was set to make sure that the objects are unresolved, hence compact, in HST imaging. Simulations show that the intrinsic physical size of these sources may around $200\,{\rm kpc}$ and it is therefore argued that these very compact sizes are indicative of the objects being low-luminosity AGNs \citep{Faisst_2022}. The color difference between HST and HSC are indicative of a strong break due to CGM absorption at $900\,{\rm \AA}$ as well as strong Ly$\alpha$ emission, which puts these candidates at $z\sim6$. Visual inspection was carried out to remove spurious detections (e.g., cosmic rays).

In 2022, we observed seven out of the 12 candidates from \citet{Faisst_2022} using HST ACS/800L grism aiming to observe their \lya~spectra. The seven target were selected by visual inspection of each of the four ACS/F814W dithers exposures to be the most robust candidates out of the 12. Specifically, it is required that their brightness is constant within photometric uncertainties in all the frames to reject cosmic rays or other artificial systematics.
%In 2022, we observe 7 targets from \citet{Faisst_2022} using HST ACS/800L grism aiming to observe their \lya~spectra. The seven targets are selected by visual inspection of each of the four F814W exposures to remove spurious objects (e.g. cosmic rays).
Confirming the \lya~spectra can be used to confirm the redshifts of the objects, and to determine the AGN luminosity function of \muv$\sim-21$ at $z=6$.

This paper is structured as follows.  In Section~\ref{sec:data} we present the follow-up observation using the HST ACS grism for 7 objects, and summarize the accessible data for our sample including 1 objects that are covered by the COSMOS-Web JWST imaging.  In Section~\ref{sec:analysis}, we construct models to explain the observational constraints.  In Section~\ref{sec:discussion} we discuss the AGN luminosity function, the comparison-sample to the LRDs, and the contamination in our sample. 
Throughout this work, we assume a $\Lambda$CDM cosmology with $H_0=$70km s$^{-1}$ Mpc$^{-1}$, $\Omega_\Lambda=$0.7, and $\Omega_m=$0.3. All magnitudes are given in the AB system.

%%%%%%%%%%%%%%%%%%%%%%%%%%%%%%%%%%%%%%%%%%%%%%%%%%%%%
%%%%%%%%%%%%%%%%%%%%%%%%%%%%%%%%%%%%%%%%%%%%%%%%%%%%%
\section{Data} \label{sec:data}
%%%%%%%%%%%%%%%%%%%%%%%%%%%%%%%%%%%%%%%%%%%%%%%%%%%%%
%%%%%%%%%%%%%%%%%%%%%%%%%%%%%%%%%%%%%%%%%%%%%%%%%%%%%
In this section, we review the available observations for our sample as well as outline the data reduction of the HST grism spectroscopy.

\subsection{Imaging Data}
%Our sample of seven candidates has been observed (RC: well, not really - you mean the field was covered) 
Our sample of seven candidates are part of the COSMOS field \citep[][]{Scoville_2007} that was covered with a wealth of ground- and space-based observatories in more than 30 different filters \citep[][]{Weaver_2022}.
%Our sample of seven galaxies was observed in more than 30 bands thanks to the wealth of observational data on the COSMOS field \citep[see][]{Weaver_2022}.
Ground-based imaging include observations from the {\it Subaru} Strategic Program (SSP) survey carried out with the Hyper-SuprimeCam (HSC) in the $i$- and $z$-band is available and taken between March 2014 to November 2015 \citep{Aihara_2018}. Of additional importance for this work are the ground-based near-infrared (NIR) filters from the UltraVISTA program with VIRCAM on the VISTA telescope \citep[$Y$-, $J$-, $H$-, and $K$-band from 2009 to 2022;][]{McCracken_2012} as well as observations with {\it Spitzer} at $3.6\,{\rm \mu m}$ (channel 1) and $4.5\,{\rm \mu m}$ (channel 2) from the {\it SPLASH} survey \citep{Steinhardt_2014}.

\begin{figure}
\includegraphics[width=0.99\linewidth]{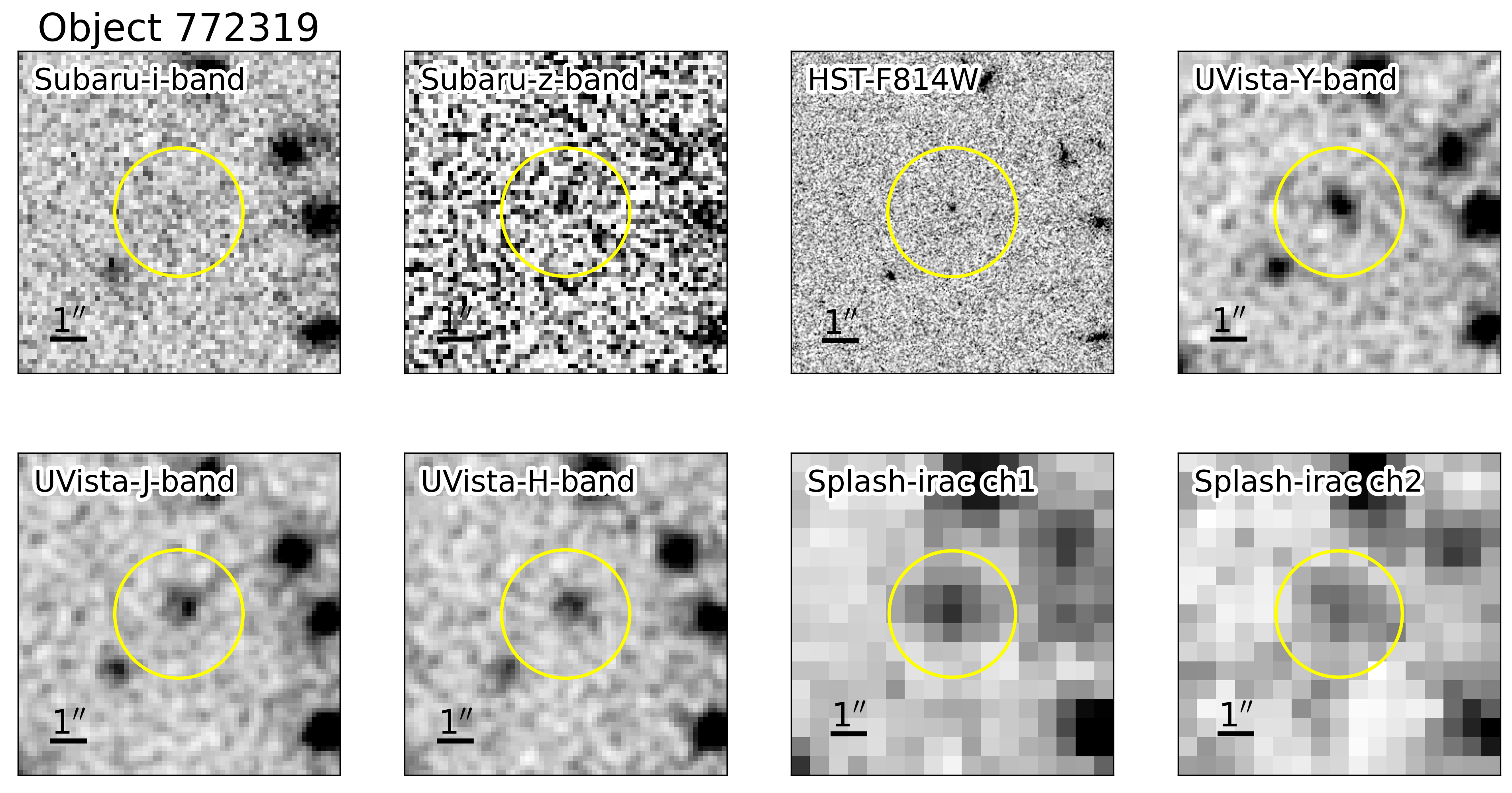}
\includegraphics[width=0.99\linewidth]{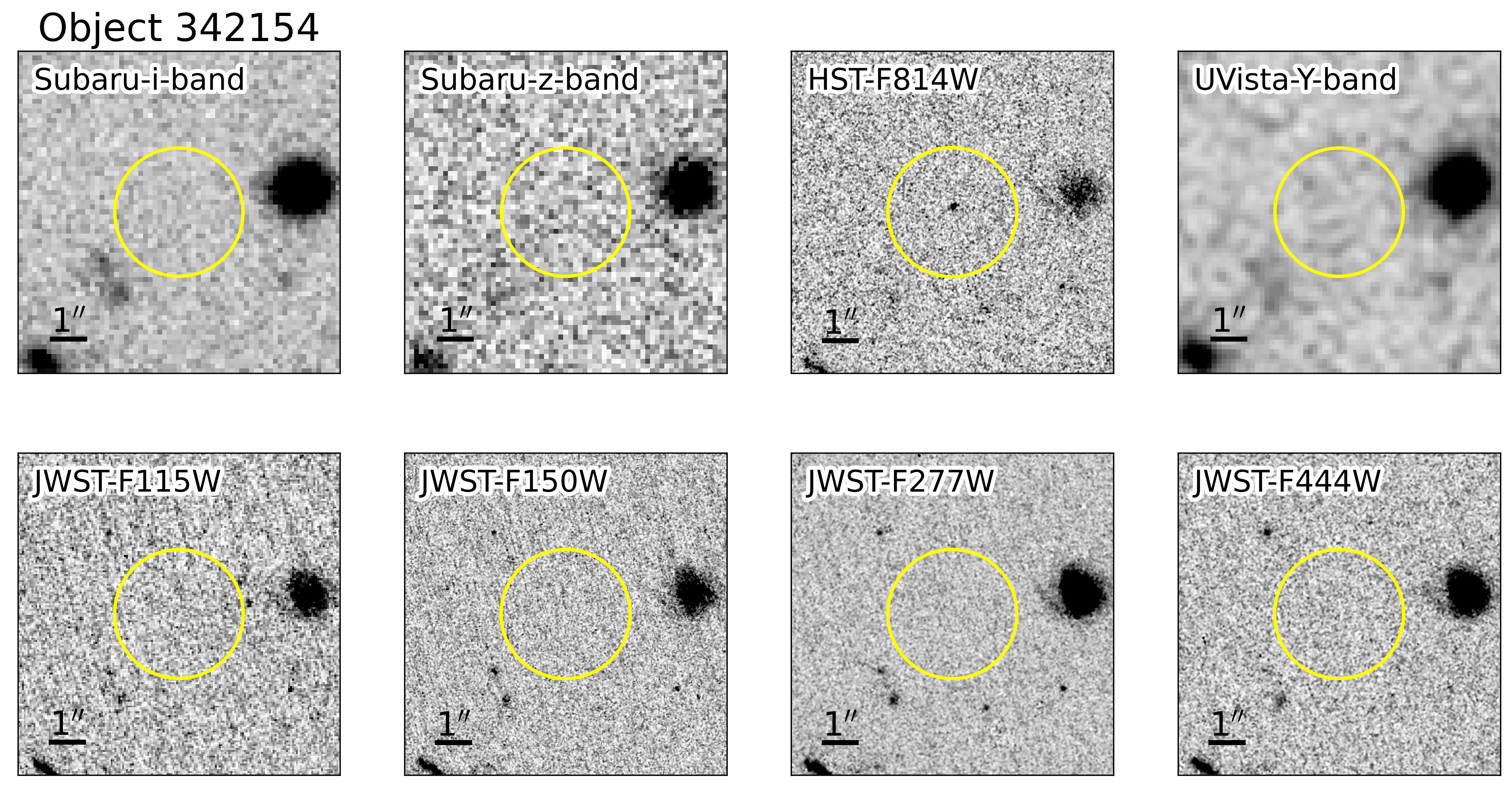}
\caption{ Top:   Images cutout of the object 772319, which is the only object in our sample detected in the multiple filters. 
Bottom: Images cutout of the object 342154 as the example of faint objects that are only detected in the F184W. The object 342154 has also been observed with JWST F115W, F150W, F277W, F444W imagings, which are 1-3 magnitude deeper than the UVista and IRAC imagings. Yet is is still undetected in the observed NIR. 
\label{fig:source_imaging}}
\end{figure}

Space-based observations include data from the {\it Hubble Space Telescope} (HST) using the ACS instrument as well as the {\it James Webb Space Telescope} (JWST) NIRCam and MIRI instruments.
All seven candidates are detected (and initially selected) in the ACS/F814W filter observed between October 15, 2003 and May 21, 2005 \citep{Koekemoer_2007, Scoville_2007}. The $5\sigma$ depth of the F814W observations are 27.20 mag \citep{Koekemoer_2007}.  One candidate (303826) is covered and marginally detected in WFC3/IR F160W from CANDELS \citep[PID:  12440, ][see Appendix~\ref{sec:appendix1}]{Grogin_2011,Koekemoer_2011}.
One object in our sample (ID 342154) is additionally covered by the {\it COSMOS-Web} cycle 1 JWST treasury program \citep[PID: 1727;][]{Casey_2023}, providing observations in the NIRCam filters F115W, F150W, F277W, and F444W as well as MIRI F770W, carried out through January 2023 to January 2024. The corresponding 5$\sigma$ depths of the {\it COSMOS-Web} NIRCam filters are 26.87, 27.14, 27.71, 27.61 magnitude, respectively (Franco et al. {\it in prep.}).
The photometry and limits are reported in table~\ref{tab:photometry}, and the image cutouts of the object 772319 and object 342154 are shown in Figure \ref{fig:source_imaging} as representations of a bright and faint source in our sample. We show the image cutouts of the remaining targets in Appendix~\ref{sec:appendix1}.

%In particular, the sample was observed (and was initially selected) in the F814W filter imaged by the {\it Hubble} Space Telescopes' Advanced Camera of Surveys (HST/ACS) between October 15, 2003 and May 21, 2005 \citep{Koekemoer_2007, Scoville_2007}. 
%Furthermore, ground-based imaging from the {\it Subaru} Strategic Program (SSP) survey carried out with the Hyper-SuprimeCam (HSC) in the $i$- and $z$-band is available and taken between March 2014 to November 2015 \citep{Aihara_2018}. 
%Of additional importance for this work are the ground-based near-infrared (NIR) filters from the UltraVISTA program with VIRCAM on the VISTA telescope \citep[$Y$-, $J$-, $H$-, and $K$-band from 2009 to 2022;][]{McCracken_2012} as well as observations with {\it Spitzer} at $3.6\,{\rm \mu m}$ (channel 1) and $4.5\,{\rm \mu m}$ (channel 2) from the {\it SPLASH} survey \citep{steinhardt_2015}.

%The sample has images observed with the Hubble Space Telescope (HST) Advanced Camera Surveys (ACS) for the F814W images through out Oct 15, 2003 to May 21st, 2005 \citep{Koekemoer_2007, Scoville_2007}; $Subaru$ Strategic Program (SSP) survey with the Subaru/Hyper-SuprimeCam (HSC) for i- and z-band through out March 2014 to November 2015 \citep{Aihara_2018}; the UltraVISTA program with the VIRCAM on the VISTA telescope for near infrared (NIR) with Y-, J-, H-, and K-band from 2009 to 2019 \citep{McCracken_2012}.  

\subsection{Grism Spectroscopy}\label{sec:data_grism}
% References and details of the grism 
Grism observations with ACS/G800L were obtained for all seven targets in seven different pointings (two orbits each) as part of the HST Cycle 30 GO proposal ID 17091 (PI: Faisst). 
The grism covers the observed wavelength from $5500 - 10,000\,{\rm \AA}$ at a resolution of $R\sim100$. For each source, two observation modes were carried out using the full-field ACS frames; one direct image ($180\,{\rm s}$) in F814W and four exposures (at $430\,{\rm s}$) in G800L at two position angles (hence in total 8 exposures) to facilitate the deblending of the dispersed light.
%The grism observation following up 7 targets was acquired with the ACS G800L grism in 7 different pointing in the HST Cycle 30 GO proposal ID 17091.  The grism covers the observed wavelength from 5500 \AA~to 10000 \AA. 
%For each sources, the grism observation include two ingredients of full-field ACS frames:  one direct image in F814W, and 8 dispersed 2D spectra in G800L (4 exposures in two position angles). 
The observation for each target was organized in two orbits split into two visits.
%For each orbit, we require a 180 second direct image in F814W. We then can fit four 430 second exposures in G800L.
We use the default dither pattern {\tt\string ACS-WFC-DITHER-BOX} for each of the observations within one orbit. This corresponds to {\tt\string POS-TARG} pairs (0\arcsec,0\arcsec), (0.247\arcsec,0.094\arcsec), (0.124\arcsec,0.232\arcsec), and (0.124\arcsec,0.138\arcsec) for optimal half-pixel sampling in both x and y, with overall dimensions large enough to help reject the larger detector artifacts. We estimate a conservative lower limit of
$2.5 \times 10^{-17}\,{\rm erg\,s^{-1}\,cm^{-2}}$ or $5.0 \times 10^{-17}\,{\rm erg\,s^{-1}\,cm^{-2}}$ for the \lya~flux if the targets are at $z=5.9$ or $6.2$, respectively. The details of the flux estimation are detailed in Section~\ref{sec:analysis}.
%We expect to detect our faintest source ($25.7\,{\rm mag}$ in $I$-band) to be detected in the F814W direct image at ${\rm SNR} \sim 3$.
Assuming a \lya~flux common for AGNs at these redshifts, we expect a detection of the \lya~line peak at signal-to-noise ratio ${\rm (SNR)>3}$ per resolution element for the faintest of our targets ($25.7\,{\rm mag}$ in $i$-band) given $3440\,{\rm s}$ of exposure \citep[see also][]{Faisst_2022}. 

%We expected to detect our faintest source ($25.72\,{\rm mag}$) in the F814W direct image at a SNR$>$3 given the 360 seconds exposure time, and detect the \lya~at a peak SNR$>$3 per resolution element for the faintest of these flux given 3440 seconds exposure time per target with the assumptions .
%Furthermore, we place the targets in the 1/3 closest to the chip gap. We do this to avoid edge-effects but at the same time to put the target as central as possible to study the surrounding environment at an even radius.

\begin{figure}
\includegraphics[width=0.95\linewidth]{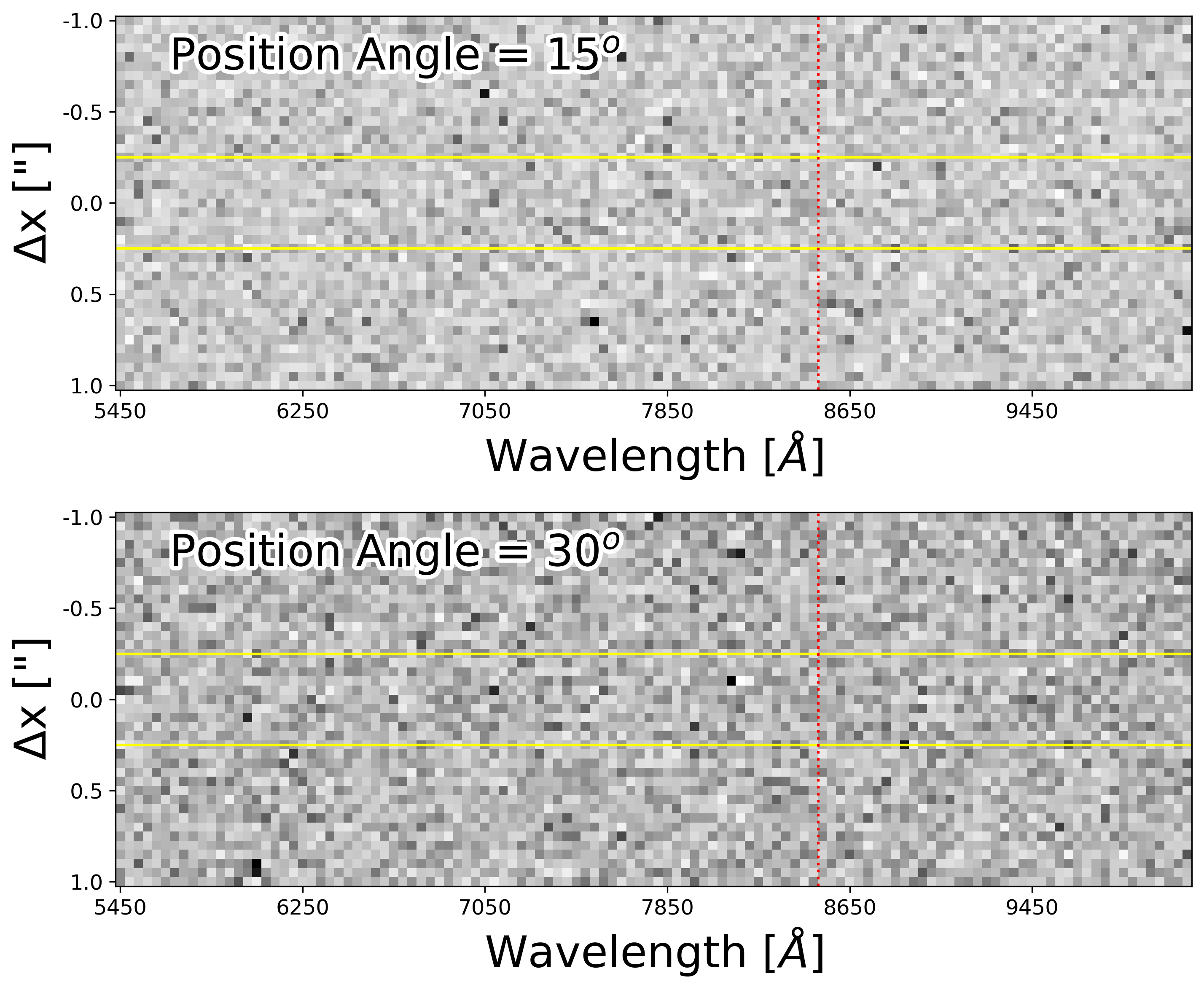}
\caption{ The grism 2D-spectra of object \galid\ at two different position angles, with each 2D spectra stacked from 4 exposures. The yellow lines mark the trace and aperture of extraction, and the red dash lines mark where the \lya~emission is expected at $z=6$. There are no common emission features in the two spectra. 
\label{fig:grism}}
\end{figure}

The grism redshift and line analysis software for space-based slitless spectroscopy \citep[\grizli;][]{Brammer_2019} was used to process, align, and coadd the exposures, as well as for the spectral extraction. 
\grizli~identifies objects from the direct images, performs a fine astrometric alignment to the Gaia Data Release 3 \citep{GAIA_2021}, and uses the direct image as the reference to establish the wavelength zero-points of the grism exposures. 
The ACS data comes from two chips, resulting to two science extensions.  Our targets are placed in the second extension approximately 1/3 closest to the chip gap, to avoid edge-effects and to ensure coverage of the full spectral spur. For \grizli~to be run correctly, the second science extension ({\tt\string sci\_extn=2}) needs to be specified in the {\tt\string GrismFLT} task in the \grizli~source code.
For spectral extraction, the standard \grizli~tasks require a given source to be detected in the direct image. However, our candidates are not detected in the direct image despite the expected ${\rm SNR \sim 3}$ computed by the {\it Hubble} exposure time calculator.
%(despite the expected ${\rm SNR \sim 3}$). {\color{blue} This could be explained by either a not accurate astrometric solution or a over-estimation of the F814W SNR value output by the {\it Hubble} exposure time calculator.} \textbf{DISCUSS POSSIBLE REASONS? IS THE SNR=3 DERIVED FROM THE ETC?}
To be able to extract the two-dimensional spectra at the location of our candidates, we therefore had to manually add mock sources (described by a Gaussian profile of the size of the F814W PSF FWHM) at the respective coordinates.
We ensure the added mock sources at pointing to the correct position in the grism spectra by checking the alignment of detected sources nearby our target to the Gaia-aligned archival F814W images. This test shows a maximal offset of 0.1\arcsec~in the astrometric solution of the direct image may be present.
We do not detect the expected \lya~emission for any of our targets in the two-dimensional spectra subsequently extracted by \grizli (See Figure~\ref{fig:grism}) . This suggests weaker \lya~emission in these galaxies compared to common values expected in high-redshift AGNs (see Section~\ref{sec:analysis}).

\begin{table*}
\centering 
\begin{tabular}{lccccccc} 
\hline\hline
 ID & 772319 & 298988   & 303826 & 342154 & 622752 & 667155 & 747548  \\
\hline
ra       &  149.4493        & 150.4077      & 150.3563 & 150.1967 & 149.7843 & 149.6293 &  149.5442   \\
dec      &  2.5234          & 2.7466        & 2.7593  & 2.0339 & 1.7169 & 2.2474 & 2.2755    \\
$i$        & $>$26.90         & $>$26.90 & 26.92$\pm$0.1 & $>$26.90 &  $>$26.90 & $>$26.90 &  27.73$\pm$0.36   \\
$z$        & 25.25$\pm$0.05   & $>$26.40      & $>$26.40 & $>$26.40 & $>$26.40 & $>$26.40 & $>$26.40    \\
F814W    & 25.72 $\pm$ 0.01 & 25.71$\pm$0.03& 25.27$\pm$0.03 & 25.41$\pm$0.02 & 25.22$\pm$0.01 & 25.63$\pm$0.02 & 25.55$\pm$0.02    \\
Y        & 25.23 $\pm$ 0.04 & $>$26.20      & $>$26.20 & $>$26.20 & $>$26.20 & $>$26.20 & $>$26.20   \\
J        & 25.02 $\pm$ 0.04 & $>$25.85      & $>$25.85 & $>$25.85 & $>$25.85 & $>$25.85 & $>$25.85   \\
H        & 24.96 $\pm$ 0.06 & $>$25.48      & $>$25.48 & $>$25.48 & $>$25.48 & $>$25.48 & $>$25.48   \\
IRAC ch1 & 23.64 $\pm$ 0.14 & $>$24.01      & $>$24.01 & $>$24.01 & $>$24.01 & $>$24.01 & $>$24.01    \\
IRAC ch2 & 24.01 $\pm$ 0.20 & $>$23.34      & $>$23.34 & $>$23.34 & $>$23.34 & $>$23.34 & $>$23.34    \\
G800L$^{a}$    &    1.5$\times10^{-19}$   &   1.5$\times10^{-19}$   &  1.5$\times10^{-19}$  &  1.5$\times10^{-19}$  & 1.5$\times10^{-19}$  & 1.5$\times10^{-19}$  &   1.5$\times10^{-19}$   \\
JWST     & --               & --            & -- & no detection & -- & -- &  --    \\
\hline
\end{tabular}
\caption{ The Object ID, photometry, and grism limits of our sample. We report the 5$\sigma$ upper limit if the source is not detected.  The Object 772319 is the only object detected in multiple bands. Object 342154 is observed with the JWST F115W, F150W, F277W, and F444W imaging, with the 5$\sigma$ depth of 26.87, 27.14, 27.71, 27.61, respectively. (a) The flux limit of the grism G800L at 8500\AA, in the unit of erg~s$^{-1}$~cm$^{-2}$.
}  
\label{tab:photometry} 
\end{table*}

%%%%%%%%%%%%%%%%%%%%%%%%%%%%%%%%%%%%%%%%%%%%%%%%%%%%%
%%%%%%%%%%%%%%%%%%%%%%%%%%%%%%%%%%%%%%%%%%%%%%%%%%%%%
\section{Constraints on Lyman-$\alpha$ Emission of the Compact AGN Candidates} \label{sec:analysis}
%%%%%%%%%%%%%%%%%%%%%%%%%%%%%%%%%%%%%%%%%%%%%%%%%%%%%
%%%%%%%%%%%%%%%%%%%%%%%%%%%%%%%%%%%%%%%%%%%%%%%%%%%%%

The seven candidate high-redshift AGNs discussed in this work are specifically selected to be at redshifts $z\sim5.5 - 6.5$ as well as compact (unresolved in HST imaging) to resemble high-redshift low-luminosity AGNs \citep[see][]{Faisst_2022}. The redshift selection was performed photometrically by requiring a red color between the {\it Subaru} $i$-band and the ACS/F814W filter. The large color difference between those filters is motivated by the fact that the Lyman break ($\lambda_0 = 912\,{\rm \AA}$) is redshifted to the $i$-band while the corresponding \lya\ emission line ($\lambda_0 = 1216\,{\rm \AA}$) and red continuum contribute to the F814W flux density.
The grism spectroscopy in combination with broad-band photometry allows us to derive a value or limit on the \lya~equivalent width (EW).
%The grism observations in ACS/G800L obtained in this work cover the redshifted \lya~emission and therefore would allow us to confirm the redshift of the candidates and to understand how much they are ionizing their surrounding intergalactic medium (IGM). 
%In addition, the grism spectroscopy in combination with broad-band photometry allows us to derive a value or limit on the \lya~equivalent width (EW).

As outlined in Section~\ref{sec:data_grism}, we do not detect any emission in the grism spectra for any of our targets, which has important implication on the nature of the candidates as well as their \lya~emission if they are at $z\sim6$, as discussed in the next sections.

The aperture for the spectral extraction is chosen to be 0.5\arcsec, which includes the possible astrometric uncertainty (estimated to be $\sim0.1\arcsec$; Section~\ref{sec:data_grism}) and is large enough to account for possible extent of \lya~emission. 
We estimate the upper limit of the \lya~EW based on the average root-mean-square (RMS) from the grism spectra, assuming the redshift of the source is between $z=5.8-7.0$ where the \lya\ emission is redshifted into the F814W filter but not contaminate the {\it Subaru} HSC $i$-band.
Note that at the wavelength resolution of the grism spectra ($R\sim100$, corresponding to $\sim50\,{\rm \AA}$ or $1,800\,{\rm km\,s^{-1}}$ for \lya\ line at $z\sim6$), the \lya~emission is expected to be contained in one or two spectral resolution elements. 
We assume in the following that the source is described by a UV continuum in the form of a power law ($f_\lambda \propto \lambda^\beta$, where $\beta$ is also referred to as the UV continuum slope) superimposed by a \lya~emission line at rest-frame $1216\,{\rm \AA}$. 
The \lya~flux is parameterized as a function of a given SNR observed value
\begin{equation}\label{eq:lya}
{\rm SNR} \times \sigma_g = F(Ly\alpha) + f_{\rm UV,cont} \times {\rm d}\lambda, 
\end{equation}
where $f_{\rm UV,cont}$ is the underlying continuum flux density, $\sigma_g$ is the uncertainty per spectral element, and d$\lambda$ is the rest-frame width of the spectral element. 
Given the depth of the grism spectrum, we derive the 3$\sigma$ upper limit of $F$(\lya) $\leq 2.3\times10^{-17}$ erg~s$^{-1}$~cm$^{-2}$.
We vary the observed redshift of the spectrum from $z = 5.8 $ to $z = 7$ and normalize it in the F814W filter to the observed magnitude of the candidates. 

\begin{figure}
\includegraphics[width=0.99\linewidth]{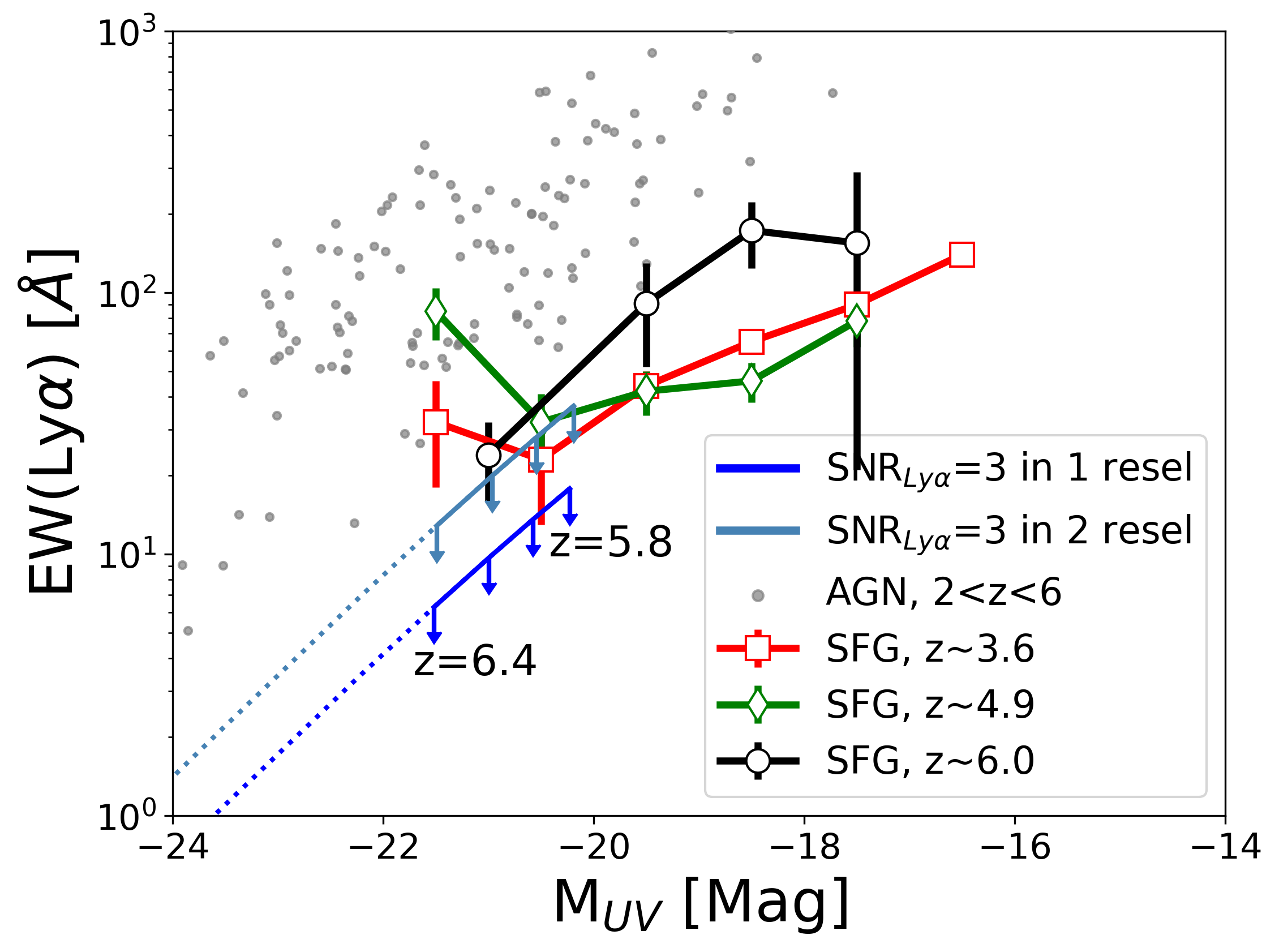}
\caption{  The rest-frame EW(\lya) vs the UV magnitude (\muv). We derive the EW(\lya) that are consistent to the F814W flux density and mark the \lya\ flux SNR=3 in one resel (blue) and in two resels (steelblue) based on the grism spectrum noise level.  We compare the derived EW(\lya) upper limit to the AGN \cite[gray dots;][]{Sobral_2018} and SFG \citep[red, green, and black lines;][]{Hashimoto_2017}.
\label{fig:LyaEW_lim}}
\end{figure}

We show the upper limit of EW(\lya) in  Figure~\ref{fig:LyaEW_lim}. We also show different observed EW(\lya)$-$\muv~relations of star forming galaxies \citep{DeBarros_2017, Hashimoto_2017} as well as AGN \citep{Sobral_2018} at $2<z<6$ in Figure~\ref{fig:LyaEW_lim}. 
It is to note that there is seemingly no strong redshift evolution of this relation between $z=3.6-6.0$ for star forming galaxies.

We find that the EW(\lya) obtained by our simulations for the seven candidates are below the observations for both star forming galaxies and AGN at similar redshifts. 
Comparing to the \lya~emitters of similar \muv at lower redshift, it is therefore likely that these seven AGN candidates are either not \lya~emitters or that they are unable to ionize the surrounding neutral hydrogen during the EoR.

%%%%%%%%%%%%%%%%%%%%%%%%%%%%%%%%%%%%%%%%%%%%%%%%%%%%%
%%%%%%%%%%%%%%%%%%%%%%%%%%%%%%%%%%%%%%%%%%%%%%%%%%%%%
\section{Discussion} \label{sec:discussion}
%%%%%%%%%%%%%%%%%%%%%%%%%%%%%%%%%%%%%%%%%%%%%%%%%%%%%
%%%%%%%%%%%%%%%%%%%%%%%%%%%%%%%%%%%%%%%%%%%%%%%%%%%%%

\begin{figure}
\includegraphics[width=0.99\linewidth]{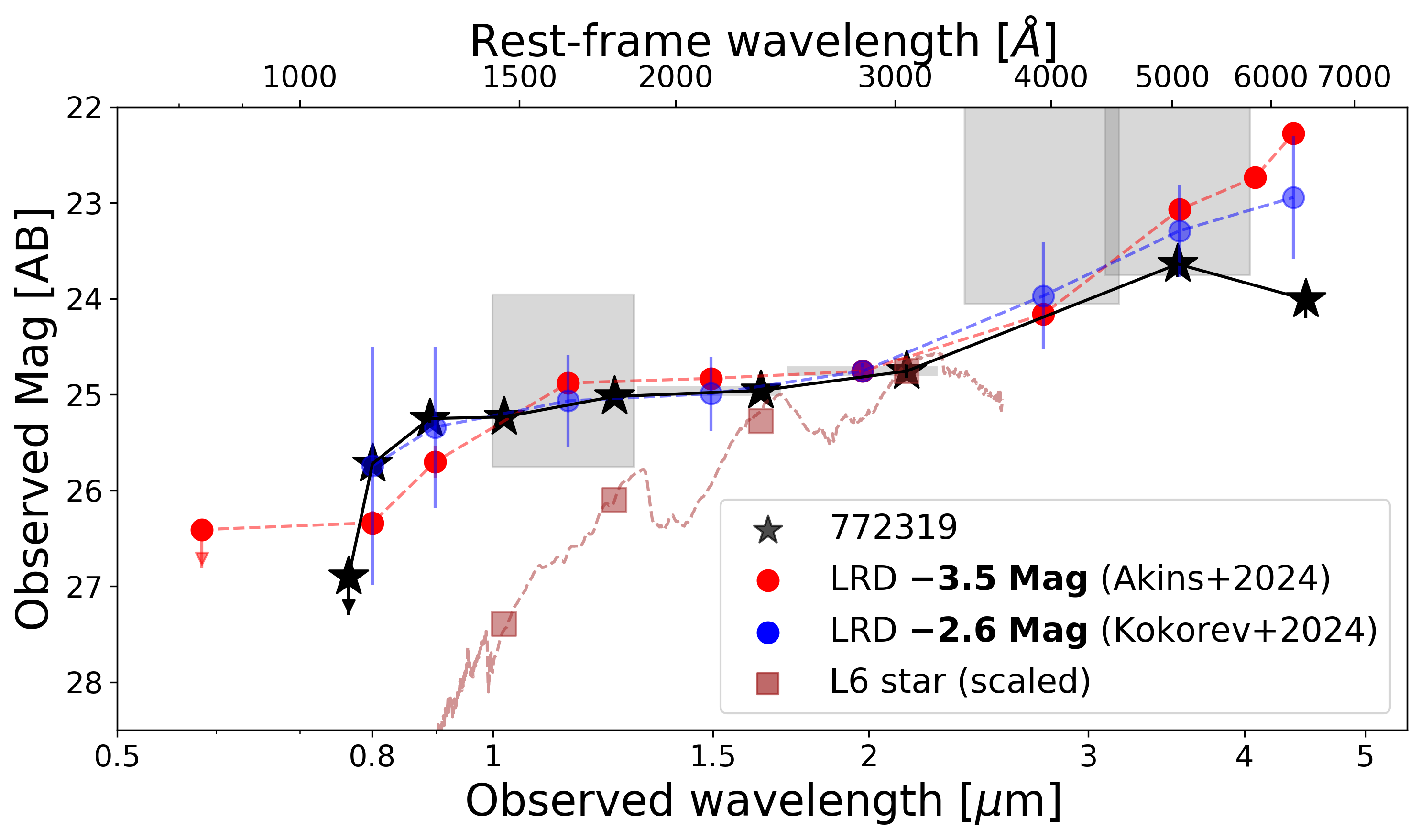}
\caption{  The SED of \galid, comparing with the rescaled SED of the spectrum of a cool dwarf from \citet{Looper_2008} (brown), and the average LRDs in \citet{Akins_2024} (Red), and \citet{Kokorev_2024} (Blue), normalized at 2$\mu$m. Notice that \galid\ is 3.5 magnitude and 2.6 magnitude brighter than the JWST selected LRDs from in \citet{Akins_2024}, and \citet{Kokorev_2024}, respectively. The color selections of the LRDs from \citet{Kokorev_2024} are shown as the gray area.   \label{fig:LRD-SEC_compare}}
\end{figure}

%%%%%%%%%%%%%%%%%%%%%%%%%%%%%%%%%%%%%%%%%%%%%%%%%%%%%
\subsection{ Brightest Little Red Dots  }\label{sec:LRD_SED}
%%%%%%%%%%%%%%%%%%%%%%%%%%%%%%%%%%%%%%%%%%%%%%%%%%%%%

Candidate source \galid~is the only one in our sample that is detected in multiple bands in the near infrared (Figure~\ref{fig:source_imaging}, top panel). An SED fit to its broad-band photometry suggests $z_{\rm phot} = 5.9$ \citep{Weaver_2022}.
Its UV magnitude of \muv$\sim-21$ puts it at the faint end of the quasar LF based on sources observed prior to the JWST era, whilst at the same time at the bright end of the faint AGN candidates discovered by JWST (including the LRDs).
In the following, we compare the SED of candidate \galid~to JWST-selected AGN. Motivated by the compactness of the source \citep[$R_e<0.06$\arcsec,][]{Akins_2024}, we show in Figure~\ref{fig:LRD-SEC_compare} the SED of \galid~(black) together with the stacked/averaged SEDs of JWST-selected LRDs from \citet{Kokorev_2024} (blue) and \citet{Akins_2024} (red). We also compare these AGN candidates with the spectrum of a cool dwarf from \citet{Looper_2008}. The SED of cool dwarfs are distinct from the AGN candidates, as the cool dwarf has the NIR light much brighter than its optical light.  
We illustrate the color selection criteria  of low redshift ($z<6$) LRDs (grey area) from \citet{Kokorev_2024}: 
\begin{eqnarray*}
F115W - F150W&<&0.8,\\
F200W - F277W&>&0.7,\\
F200W - F356W&>&1.0.
\end{eqnarray*}
We note that \galid~would be selected as LRD based on its SED as part of the color selection function applied by \citet{Kokorev_2024} (indicated by the gray shaded area).\footnote{We note that \galid~has not been covered by JWST. However, it is detected in {\it Spitzer} filters, which are used here as proxies for NIRCam F356W and F444W.}
However, although \galid~is very similar to the median SED shape of JWST-selected LRDs, it is on average 2.6 (3.5) magnitudes {\it brighter} compared to the Kokorev et al. (Akins et al.) selected LRDs. Since this magnitude different is present in the rest-frame optical (at $0.4-0.7\,{\rm \mu m}$), this indicates that \galid~is approximately $10-30$ times more massive than the LRDs shown here. 
Furthermore, \galid~has a bluer observed {\it Spitzer} $[4.5\,{\rm \mu m}]$ - $[3.6\,{\rm \mu m}]$ color compared to the LRDs (measured from F444W-F356W photometry). Since H$\alpha$ is covered by {\it Spitzer} channel 2 and NIRCam/F444W at $z = 6$, a bluer color may indicate a {\it weaker} H$\alpha$ emission in \galid. This would be consistent with the order of magnitude higher stellar mass of this source compared to the LRDs, assuming a decline in H$\alpha$ strengths (specifically EW) or reduced star formation burstiness at higher stellar masses \citep[c.f.][]{Faisst_2019}.
Notably, the luminosity of \galid~makes it the third brightest and likely one of the most massive LRD among the 300 JWST-selected LRDs in the {\it UNCOVER} and {\it COSMOS-Web} surveys with \muv$<$-21. It could therefore be an important case study of a matured LRD at the end of the EoR. 
%Importantly, the fact that its EW(\lya) is lower than the average value of AGN and star forming galaxies at similar redshifts and UV magnitudes, suggests that it is not residing, currently, in an ionized bubble, or that its intrinsic \lya~is weak compared to the cohort.
Importantly, the fact that the observed EW(\lya) is lower than the average value of AGN and star forming galaxies at similar redshifts and UV magnitudes, suggests that this object may have weaker apparent \lya~compared to its cohort (either due to intrinsically faint \lya, or stronger dust attenuation), or that is not residing, currently, in an ionized bubble.

%We find that the out line of \galid\ SED fits in the selection criteria of the little red dots presented in \citet{Kokorev_2024} (gray shaded area in Figure~\ref{fig:LRD-SEC_compare}).  
%We show in Figure~\ref{fig:LRD-SEC_compare} that the source \galid\ is very similar in SED shape to the median SED of the LRDs in \citet{Kokorev_2024} an the stacked LRDs in \citet{Akins_2024}, but are 2.6 Magnitude and 3.5 Magnitude brighter, respectively.  
%Object 772319 has are redder at observed  4.4$\mu$m, which covers the \ha\ emission if the object is at $z\sim6$.  The red color between F356W and F444W can be found in the LRDs sample from \citet{Kokorev_2024, Akins_2024}, and brighter LRDs tend to have redder color (larger magnitude in F356W$-$F444W). 
%If we calculate the luminosity of the LRDs with the photometric redshifts, \galid\ become the 3rd brightest LRDs among the 300 LRDs selected in the UNCOVER survey and the COSMOS-Web survey,  with \muv$<$-21. 

\begin{figure}
\includegraphics[width=0.99\linewidth]{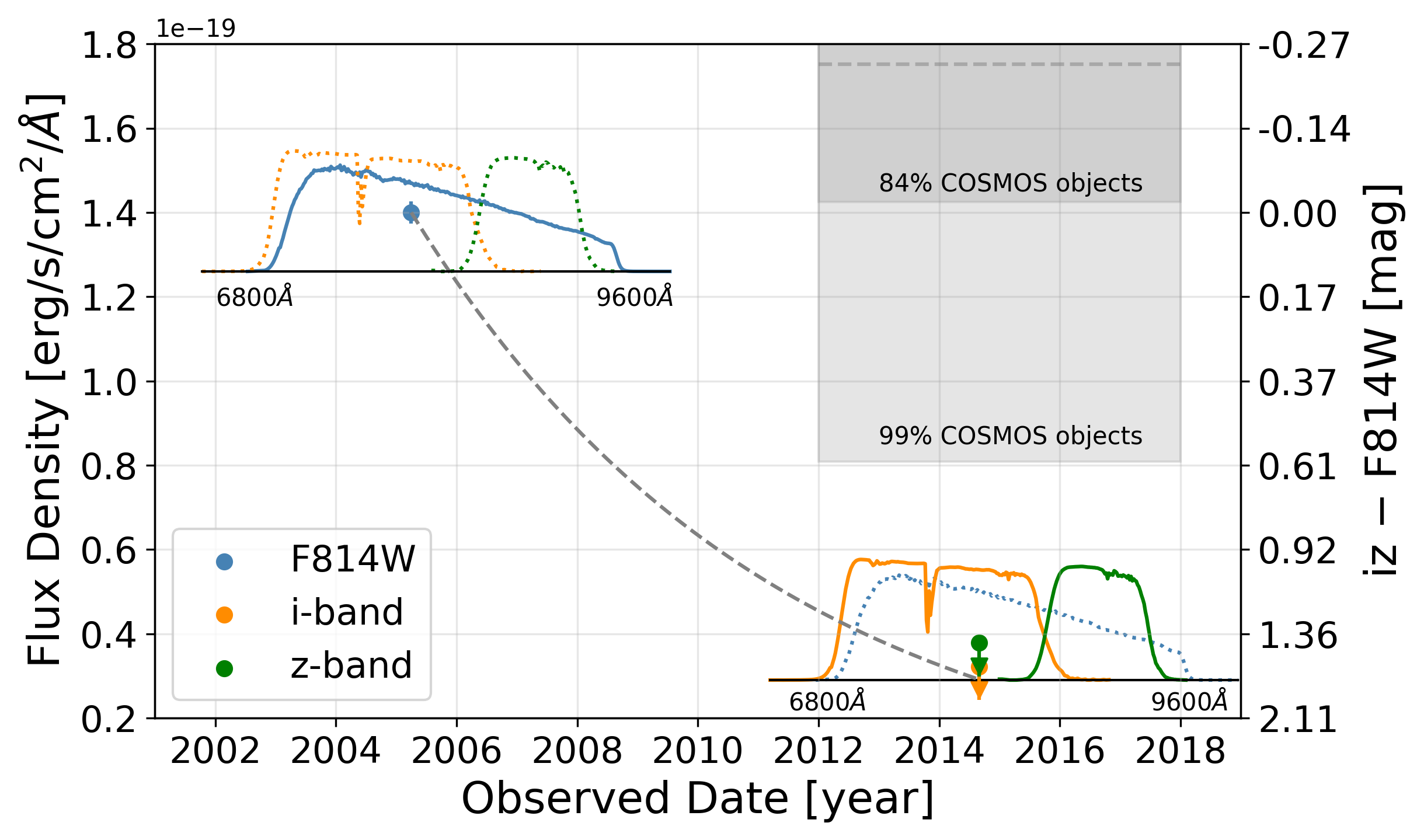}
\caption{  The iz$-$F814W colors vs the observation time.  The transmission curve of the F814W, $i$-, and $z$-band filters are illustrated on the two observations. On 2003, the sources are detected with the F814W filter, while on 2013, the sources fluxes in the similar wavelength covered by Subaru $i$- and $z$-band were undetected.  The gray dash line marks the median $iz-$F814W$\,=\,$0.24 in COSMOS2020 catalogue \citep{Weaver_2022}. \label{fig:light_curve}}
\end{figure}

%%%%%%%%%%%%%%%%%%%%%%%%%%%%%%%%%%%%%%%%%%%%%%%%%%%%%
\subsection{Possibility of transient point sources}\label{sec:transient}
%%%%%%%%%%%%%%%%%%%%%%%%%%%%%%%%%%%%%%%%%%%%%%%%%%%%%
% A bit more introduction of rejecting the star scenario (take about other bands)
Apart from \galid, the rest of the objects in the sample are only detected in F814W (but {\it nota bene} in all $4$ frames taken within one visit), but not in any other bands nor in grism spectroscopy. Intriguingly, candidate 342154 is also not detected even in JWST NIRCam/F115W imaging at a depth of $26.9\,{\rm mag}$.
In order for galaxies or AGN to be consistent with such photometric constraints, they would have to be characterize by an extremely blue dust-free UV continuum with $\beta = -6$. Such blue slope is impossible to reconcile even with a single young dust-free blue stellar population and $100\%$ continuum escape fraction.
The explanation of these objects being cool dwarfs can be excluded securely as well, since we would expect such stars to be detected in the most near-infrared filters (specifically JWST F115W, F150W and UltraVISTA $Y$ and $J$ bands), even in the unlucky case of stellar atmospheric absorption line coinciding with some of these bands.
In this section, we instead explore another explanation, namely a transient origin. Such an explanation may be reasonable given the different observation times of the different filters in which these sources are detected or not detected.
%In Figure~\ref{fig:source_imaging} bottom panel we show that the object 342154 is not detected even in JWST F115W as deep as 26.87 magnitude. 

The filter F814W covers roughly the same range as the combination of the $i$- and $z$-band filters, as shown in Figure~\ref{fig:light_curve}. Therefore, for a given spectrum, the flux density detected in F814W should be consistent with the averaged flux density in $i$- and $z$-band; if a source is detected in F814W but not in $i$-band, then the flux in F814W band should all origin from the $z$-band. In this case, if all the light detected in F814W is distributed in the $z$-band, this would result in a $z$-band observed magnitude of $>24.61$, which is $5\sigma$ above the $z$-band sensitivity limit of $26.40$.
%Vice versa, if all the light detected in F814W is distributed in $z$-band, then the $z$-band should be as bright as \textbf{XXX} magnitudes.
However, the all remaining six sources are not detected in either $i$- nor $z$-bands.

Two possible scenarios may explain these observations in F814W, $i$-, and $z$-band.
The first explanation is that the flux density of F814W is contributed from an emission line concentrated at wavelength of the small gap between $z$-band and F814W.
However, this implies that the emission line flux would be as bright as 4.6$\times10^{-16}$ erg~s$^{-1}$~cm$^{-2}$, which is in tension with the grism constraints on the line emission calculation in Section~\ref{sec:analysis}.  %Likewise, if object 342154 is a high redshift ($z\sim6$) compact galaxy,  then the UV continuum slope $\beta<-6$ would be unnaturally steep since the object is not detected in JWST F115W and F150W.
The second possible explanation would involve temporal changes in the luminosity of the sources between the observations. Since the F814W images were taken in 2003, and the Subaru $i$- and $z$- band images are taken in 2013, our sources may be variable point sources that are brighter in 2003 and fade out 10 years later (in observed frame) in 2013.

To investigate this further, we compared the $iz$$-$F814W color with the sources in COSMOS2020 catalog \citep{Weaver_2022} with detected $i$-, $z$- and F814W band magnitudes (see Table~\ref{tab:photometry} for the limits). We restrict ourselves to high SNR sources by placing a cut conservative cut on the uncertainty of the flux measurements of $<0.25\,{\rm mag}$. Without applying cuts on the photometric redshift, the median $iz$$-$F814W color is $-0.24\,{\rm mag}$, with 84\% of the objects having colors $<0.02\,{\rm mag}$ and with 99\% of the objects having colors $<0.59\,{\rm mag}$.
Applying a photometric redshift cut of $z_{phot}>5$, the median $iz$$-$F814W color is $-0.20\,{\rm mag}$ with 84\% and 99\% of the objects having colors $<0.11\,{\rm mag}$ and $<0.59\,{\rm mag}$, respectively.
From this test we learn that the much larger average $iz$$-$F814W color difference of the six sources in our sample ($>1.36\,{\rm mag}$) is hard to be explained naturally by the shape of the SEDs of sources in the COSMOS field at likely any redshift.  

%\textbf{NOTE: what about these 1\% of sources in the COSMOS catalog? are there any that reach the $1.36\,{\rm mag}$ limit? I think that's what you wanted to check...}
%As for the objects with photometric redshifts $z_{photo}>$5, the median iz$-$F814W color is $-0.20$, 84\% of the objects has iz$-$F814W $<$ 0.11, and 99\% of the objects have iz$-$F814W $<$ 0.59. 
%Our sample have a much larger magnitude difference between F814W and the average of i- and z-band (iz$-$F814W), meaning that the difference is hardly due to the shape of the spectrum. 

Assuming the six sources are transients, we can describe their light curve as $L = L_0 \exp{(-T/\tau)}$, where $L_0$ is the initial luminosity in the year 2003 and $\tau$ is the decay rate in the unit of years. Applying the constraints from the current measurements, we find $\tau<6\,{\rm yr}$ in {\it observed} frame (see Figure~\ref{fig:light_curve}). We note that if the $iz$$-$F814W colors of our sample are indeed caused by a decay in luminosity, then the photometric redshifts $z_{\rm phot} \sim 6$, estimated assuming the color difference is due to the Lyman break, are obviously no longer valid. The transient event can be at any redshift that is distant enough to be faint and unresolved.

We also investigated the case in which our sources are asteroids in our solar system. However, we rule out this case due to the fact that the centroids of the sources do not shift by more than one pixel ($0.05^{\prime\prime}\,{\rm px^{-1}}$) across all exposures in the HST ACS/F814W observations -- $600\,{\rm s}$ long in successive order \citep{Scoville_2007,Koekemoer_2007}. For example, Main Belt asteroids would move on average speeds of 1 pixel per minute. Kuiper Belt asteroids, which represent the slower end of asteroid motion, typically move at about $0.2\,{\rm mas\,s^{-1}}$, in which case they would appear approximately 4 pixels apart between exposures. Asteroids would therefore be detectable by moving across the frames.
Likewise, we rule out the possibility that the missing objects are faint galactic dwarf stars that moved due to proper motion. The proper motion of such stars is typically small \citep[$<10\,{\rm mas\,yr^{-1}}$][]{GAIA_2021}, making it unlikely that they would have shifted beyond our aperture.  

We also consider whether the color difference could be attributed to AGN variability. However, the observed magnitude difference of $>$1.36 is significantly larger than the known AGN variations in the rest-frame UV \citep{Welsh_2011} and optical \citep{Vanden_Berk_2004, Poulain_2020}. It also exceeds the variability seen in other LRDs \citep{Zhang_2024}. Major AGN flares and changing-look AGNs \citep{MacLeod_2016, Graham_2017} can produce such large magnitude differences. However, it is unlikely to explain all our missing objects with these extreme AGN, as the changing events are rare \citep[51 events from over 900,000 AGNs][]{Graham_2017} and are unlikely to outnumber the AGNs themselves. 

Our missing objects can also be explained by the contamination of supernova explosions (SNe) or tidal disruption events that occurred in host galaxies that are undetected. Though we are unable to constrain the possible redshift of the SNe, we estimate the hostless SNe rate as the following: 
We calculate the incomplete stellar mass by comparing the mass distribution of F814W-detected galaxies in COSMOS with the Schechter functions derived in \citet{Weaver_2023}. 
By integrating the incomplete galaxies mass down to 10$^6$ M$_\odot$ from $z = 0-6$, we find 3$\%$ of the stellar mass, mostly the low mass galaxies in higher redshifts, are not detected.  
\citet{Yasuda_2019} reported 1824 SNe in COSMOS throughout a half-year observation.  
Assuming the SNe rate is the same in galaxies with different stellar masses, the hostless SNe rate is estimated to be 55 degree$^{-2}$~yr$^{-1}$.  Therefore, our 6 missing objects in the 2 degree$^2$ can be a fainting hostless SNe.

\begin{figure}
\includegraphics[width=0.99\linewidth]{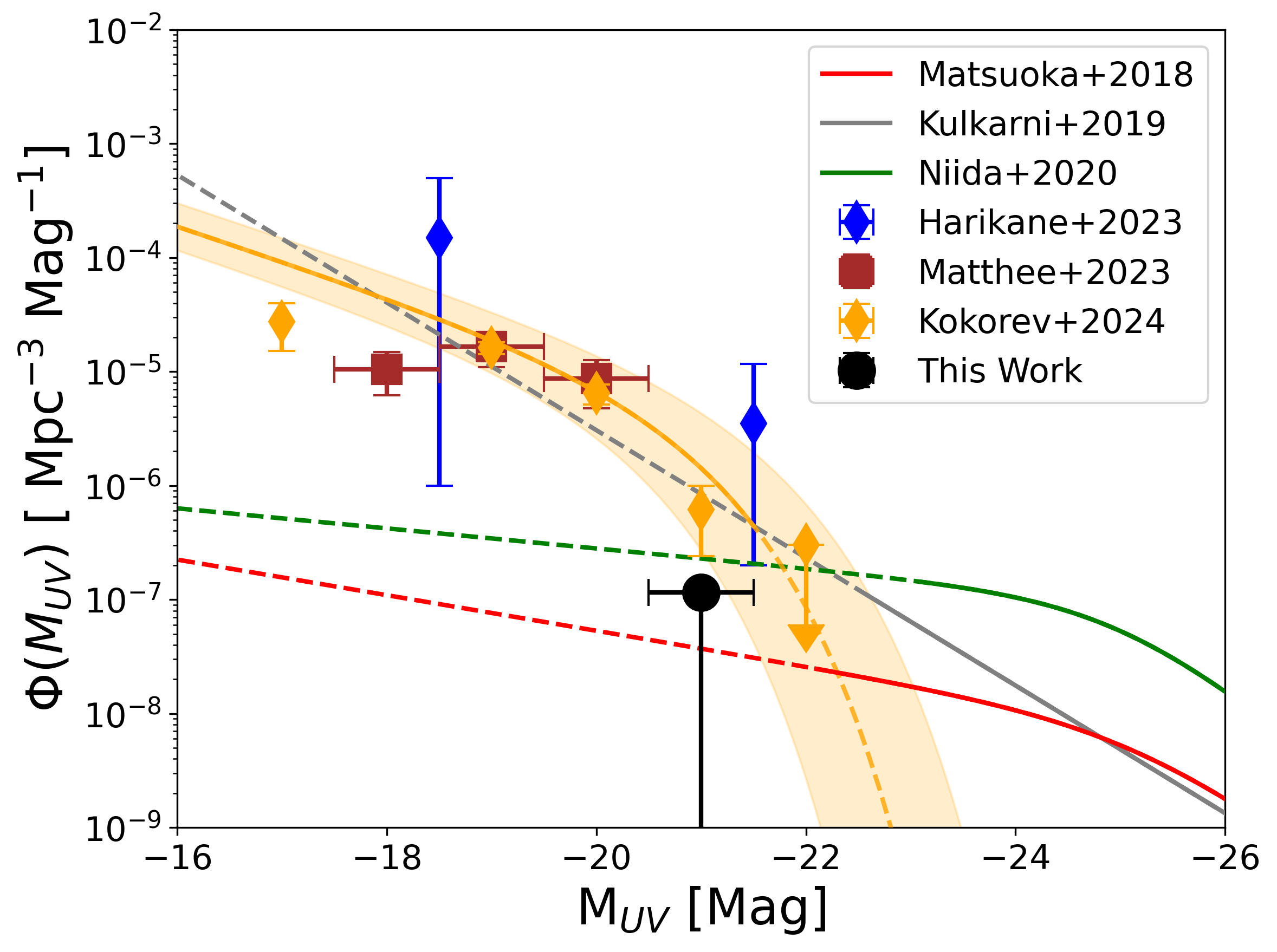}
\caption{ The AGN UV luminosity function at redshift $z=6$. We place the AGN density of  $1.1\times 10^{-7}$ Mpc$^3$ mag$-1$ as an upper limit at $z = 6$. We compare our result (black dot) with the double-power law luminosity functions derived from brighter AGNs in \citet{Matsuoka_2018} (red line), \citet{Kulkarni_2019} (gray line), and \citet{Niida_2020} (green line), where we mark solid lines are fitted from data, and dash lines are the extrapolations. We also show the faint AGNs discovered with JWST, including broad-H$\alpha$ obejcts in \citet{Harikane_2023, Matthee_2024}, and the LRDs selected from \citet{Kokorev_2024} (orange diamond). The orange line and shadows are the luminosity function of the LRDs derived in \citet{Kokorev_2024}. 
\label{fig:AGN_LF}}
\end{figure}

%%%%%%%%%%%%%%%%%%%%%%%%%%%%%%%%%%%%%%%%%%%%%%%%%%%%%
\subsection{ The AGN luminosity function of \muv$\,=\,-21$ at redshift z = 6.  }
%%%%%%%%%%%%%%%%%%%%%%%%%%%%%%%%%%%%%%%%%%%%%%%%%%%%%
Given the information on one secure detection in multiple band of our candidates and the possibility of transients for the remaining six sources, we are now able to place upper limits on the AGN LF at $z=6$ at \muv$=-21$. As mentioned before, our candidates reside at a luminosity connecting the samples of JWST-selected faint AGN (and LRDs) with bright quasars selected by other surveys not using JWST data.

\citet{Faisst_2022} select seven faint $z=6$ AGN candidates from the 2 deg$^2$ COSMOS field based on their compact size ($r_e < 0.1^{\prime\prime}$), smaller than the average galaxy at similar redshift and luminosity ($R_e > 0.2^{\prime\prime}$), measured on F814W images, as well as the color difference between F814W and the {\it Subaru} HSC $i$-band. Spurious sources (such as cosmic rays) are removed by a visual inspection and requiring detections in all four F814W exposures taken over one visit. Only one object, \galid~with a $z_{\rm phot} = 5.9$, shows multiple detections in the observed near-infrared and is characterized by a similar SED shape as the LRDs sample that possibly host AGN (see Section~\ref{sec:LRD_SED}). The other six targets are only detected in F814W, and their origin as interloper transients is more likely than being blue and faint AGN (see Section~\ref{sec:transient}). 
 As described above, we detect no emission in the HST ACS/G800L grism for any of these candidates, placing important constraints on the \lya~EW of \galid.
%We detect no emission in the HST ACS 800L grism for these 7 targets. Only one object, 772319, shows multiple detections in the observed near infrared, and shows a similar SED shape as the LRDs sample that possibly host AGN (see Section~\ref{sec:LRD_SED}).  
%The other 6 targets are only detected in the F814W band. In Section~\ref{sec:transient} we explained that these objects are more likely to be transient objects rather than extremely blue and faint AGN.  

Given this information, we conclude that only one candidate can be considered an UV-selected AGN with \muv$=-21$ at redshift $z = 6$. In the following, we derived the number density using a $V_{\rm max}$ method \citep{Schmidt_1968}, assuming the redshift of candidate \galid~is at $z = 6$. This results in an AGN density of \muv=$-21$ at $z=6$ of $\Phi \leq 1.1\times 10^{-7}$ Mpc$^3$ mag$-1$.  
We place our AGN density to the recent quasar and AGN LFs studies in Figure~\ref{fig:AGN_LF}. 

We compare our result to the LF derived from the brighter AGN with \muv$<-22$ Mag \citep{Matsuoka_2018, Kulkarni_2019, Niida_2020}, and the faint AGN (\muv$>-22$ Mag) discovered with JWST \citep{Harikane_2023, Matthee_2024, Kokorev_2024}.  Our result at \muv$=-21$ is roughly consistent to the extrapolation of the double-power law luminosity function derived from the brighter AGN from \citet{Matsuoka_2018, Niida_2020}, while it is lower by one order of magnitude compare to the JWST-selected AGN-like objects in \citet{Harikane_2023, Kokorev_2024}. 
It is intriguing to note that despite the similarity between our object and the LRDs, the faint AGN number density we derived represens the UV-selected AGN. Therefore it is not too surprising that our result is closer to the extrapolation of the UV-selected quasar LFs. 
We note that the LF upper limit are derived only from the AGN candidates with compact morphology, assuming that the host galaxies of these AGN are too faint to be detected. Therefore, the AGN selection may be incomplete as we neglect the extended sources which may include galaxies with less dominant AGN.

\section{Conclusion} \label{sec:conclusion}
We present ACS/G800L grism follow-up observation of seven faint AGN candidates in the COSMOS field at redshift of $z=6$. The candidates were selected in \citep{Faisst_2022} by their point source-like morphology and red ACS/F814W to {\it Subaru} HSC $i$-band colors due to the Lyman break at observed $\sim8000\,{\rm \AA}$ at $z=6$. Among the sample of seven candidates, only one (source \galid) is detected by multiple bands, is estimated at $z_{\rm phot} = 5.9$, and has a similar SED shape as the so called ``little red dots'' JWST-selected faint AGN candidates. However, \galid~is $\sim3$ magnitude brighter and likely $10-30\times$ more massive than these LRDs. It is found bracketed by bright quasars found pre-JWST and JWST-selected faint AGN and LRDs (some being AGN). We also detect signs of a weaker H$\alpha$ EW, which may be consistent with its higher stellar mass and lower specific star formation rate.
We find an the upper limit of the luminosity density of such bright sources of $\Phi=1.1\times 10^{-7}$Mpc$^3$ mag$^{-1}$ at \muv=$-21$ at $z=6$.
Although its brightness, we cannot confirm any \lya~emission in the G800L gratings. This places a strong $5\sigma$ limit on the \lya~EW of rest-frame $20\,{\rm \AA}$, suggesting that this source is either not a \lya~emitter (e.g., low intrinsic \lya~emission) or has not ionized the surrounding intergalactic medium. 
  
The rest of the sample (only detected in F814W) shows inconsistent flux densities in F814W and {\it Subaru} HSC $i$- and $z$-bands as well as colors in $iz$$-$F814W compared to $>99\%$ of sources found in the COSMOS field. These colors would result in unphysically blue UV continuum slopes and contamination by emission lines can be excluded based on the non-detection in our ACS/G800L grism observations. Due to the $10$-year difference in the HST and {\it Subaru} observations, we argue that these sources could be transients with luminosity decay time-scales of $\tau < 6\,{\rm yr}$ in observed frame. Unfortunately, with the current data in hand, the nature of these transient sources remains elusive. %\textbf{However, if they are at $z=0$, the most likely transients would be XYZ.} 

\section{Acknowledgement}
All the {\it HST}  data used in this paper can be found in MAST: \dataset[10.17909/dh5x-d305]{http://dx.doi.org/10.17909/dh5x-d305}.
All the {\it COSMOS} imaging data used in this paper can be found in IRSA \citet{cosmos_hst}, and the COSMOS2020 catalog can be found in \citet{cosmos2020}. %\dataset[10.26131/IRSA563]{http://dx.doi.org/10.26131/IRSA563}.
The figures of 2D grism spectra are available on Zenodo: \dataset[10.5281/zenodo.15171451]{http://dx.doi.org/10.5281/zenodo.15171451}.  The spectra underlying this study are available from the corresponding author upon request.

\clearpage
\appendix 

\section{Image Cutout of our targets}\label{sec:appendix1}

\begin{figure}[h!]
\includegraphics[width=0.47\linewidth]{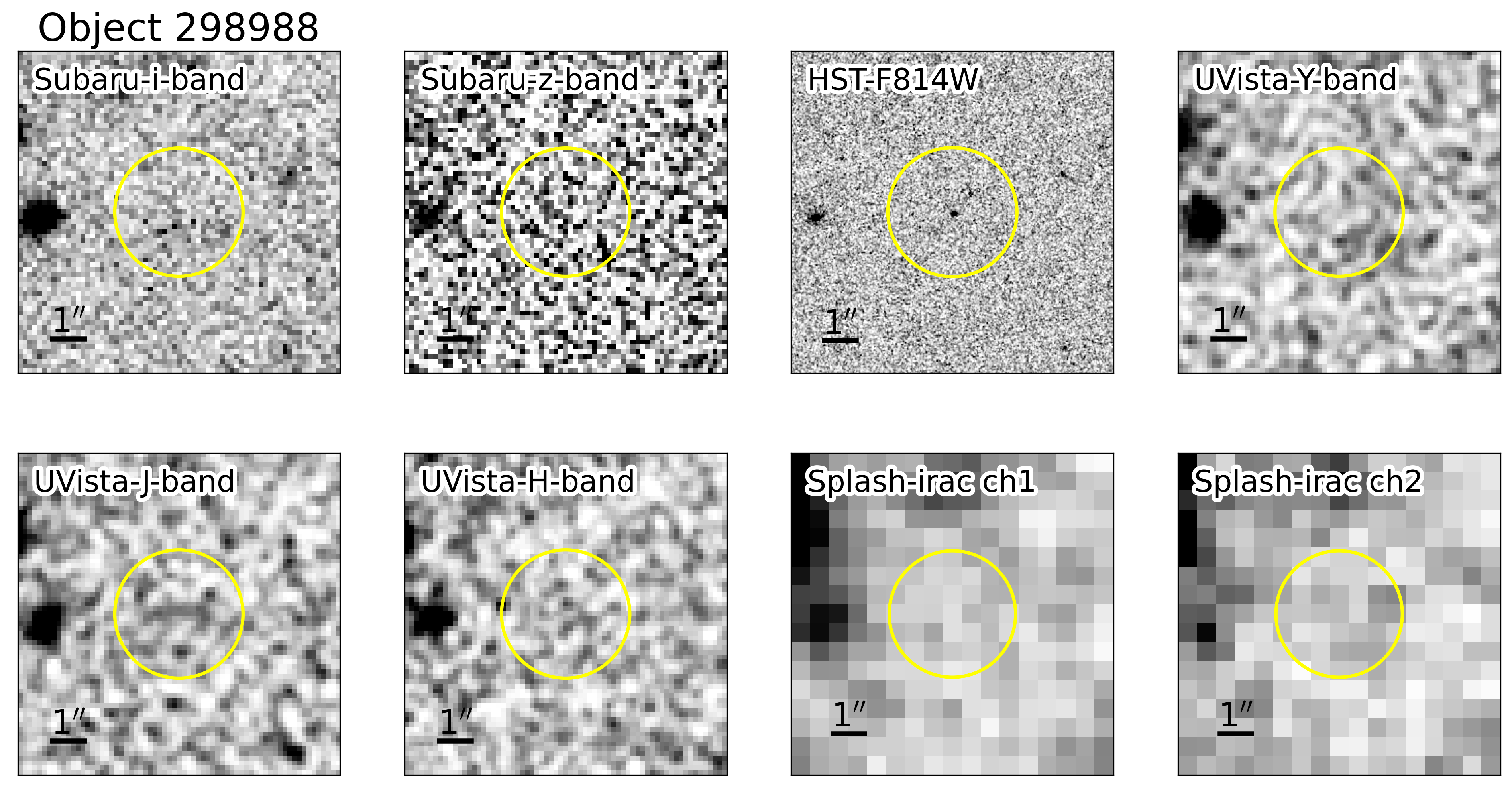}
\includegraphics[width=0.47\linewidth]{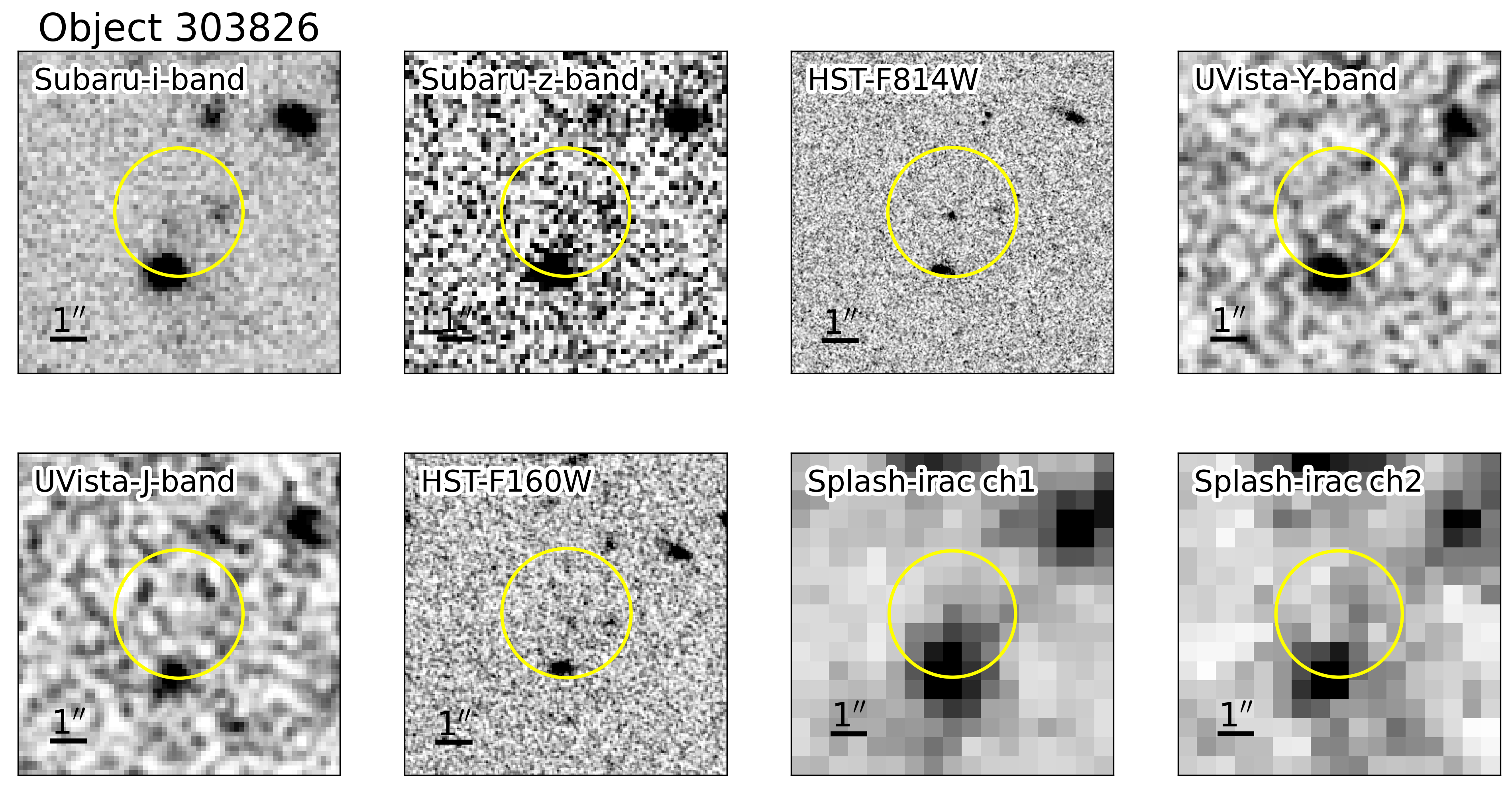}
\includegraphics[width=0.47\linewidth]{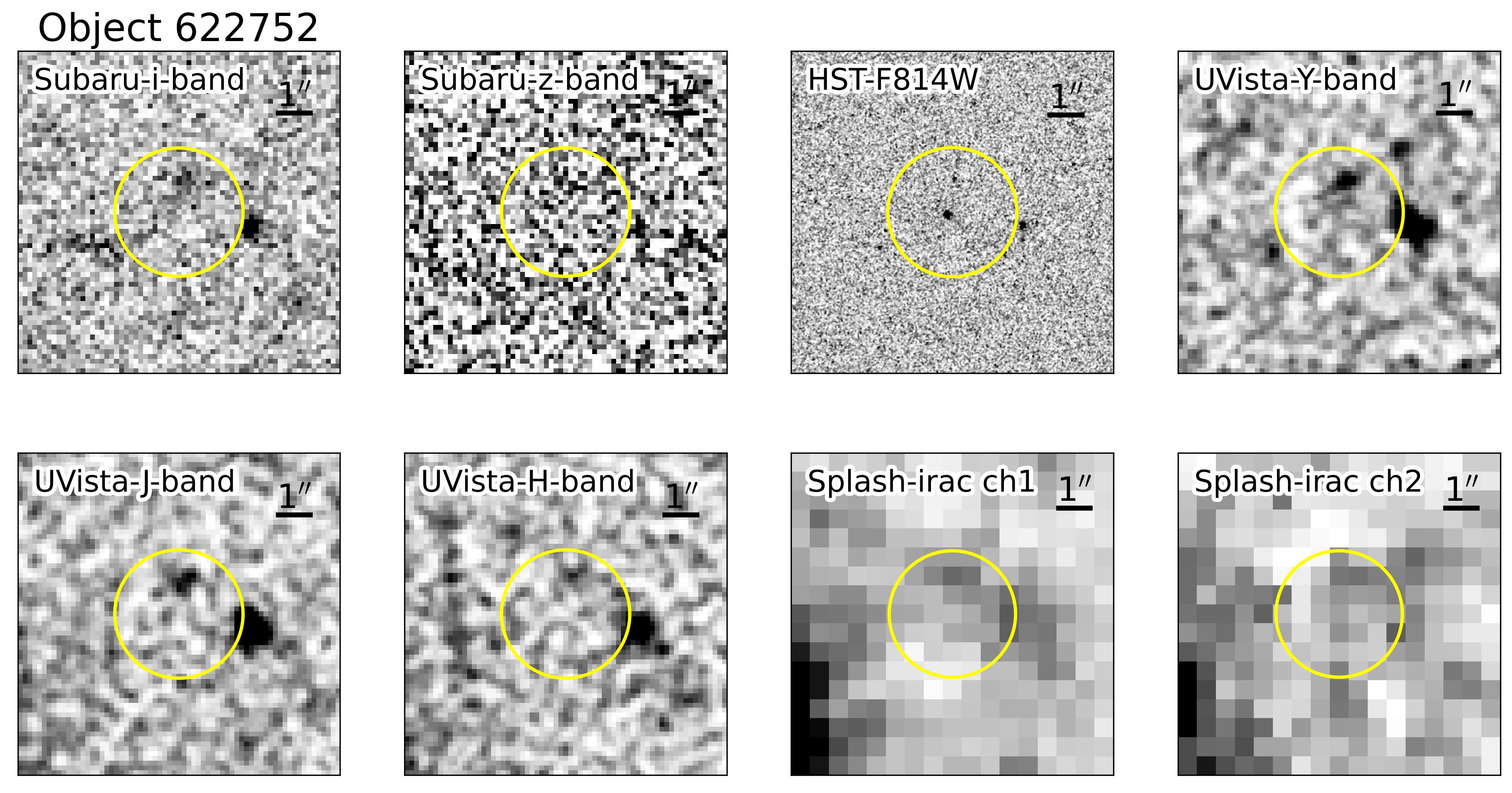}
\includegraphics[width=0.47\linewidth]{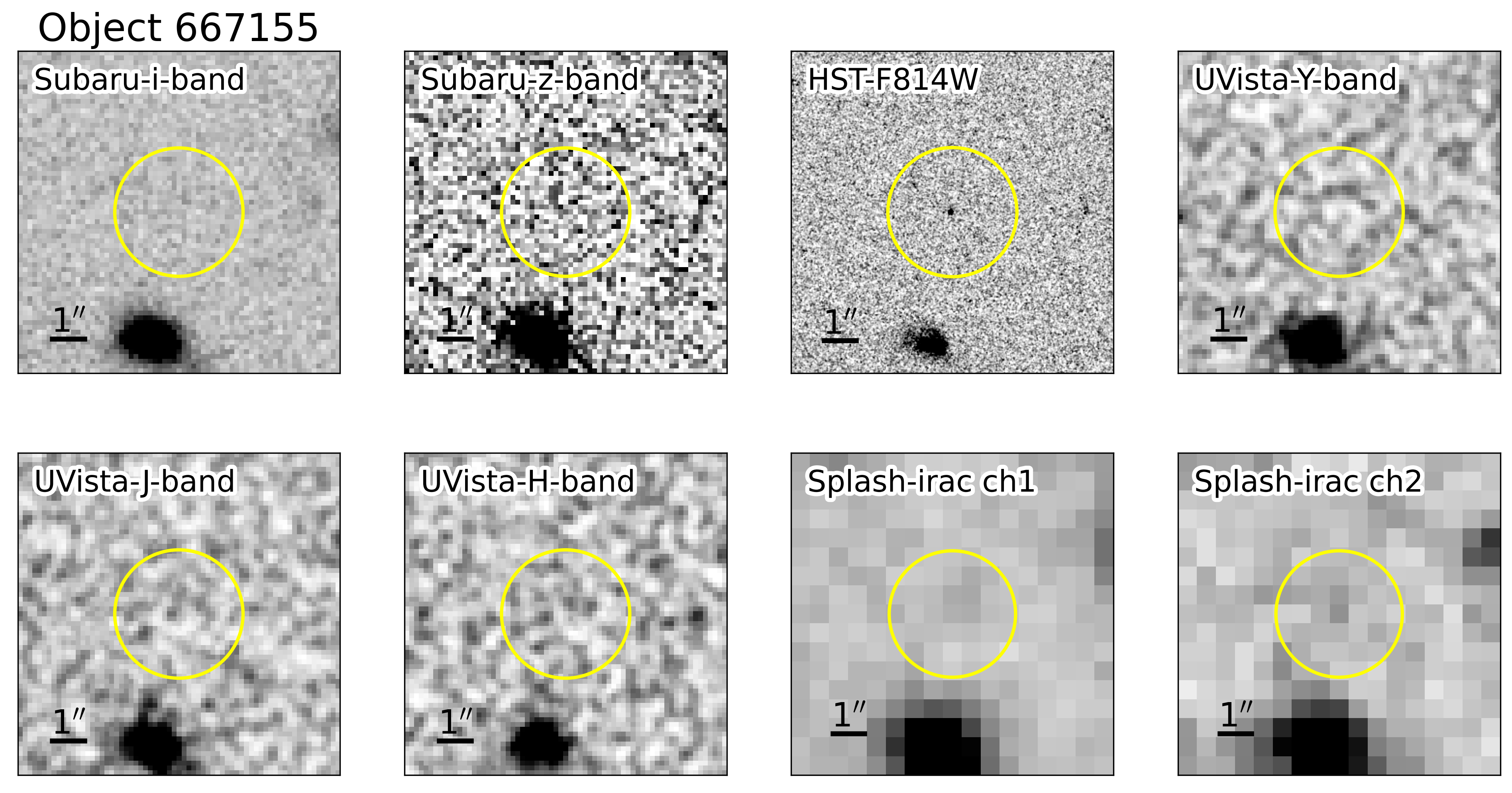}
\includegraphics[width=0.47\linewidth]{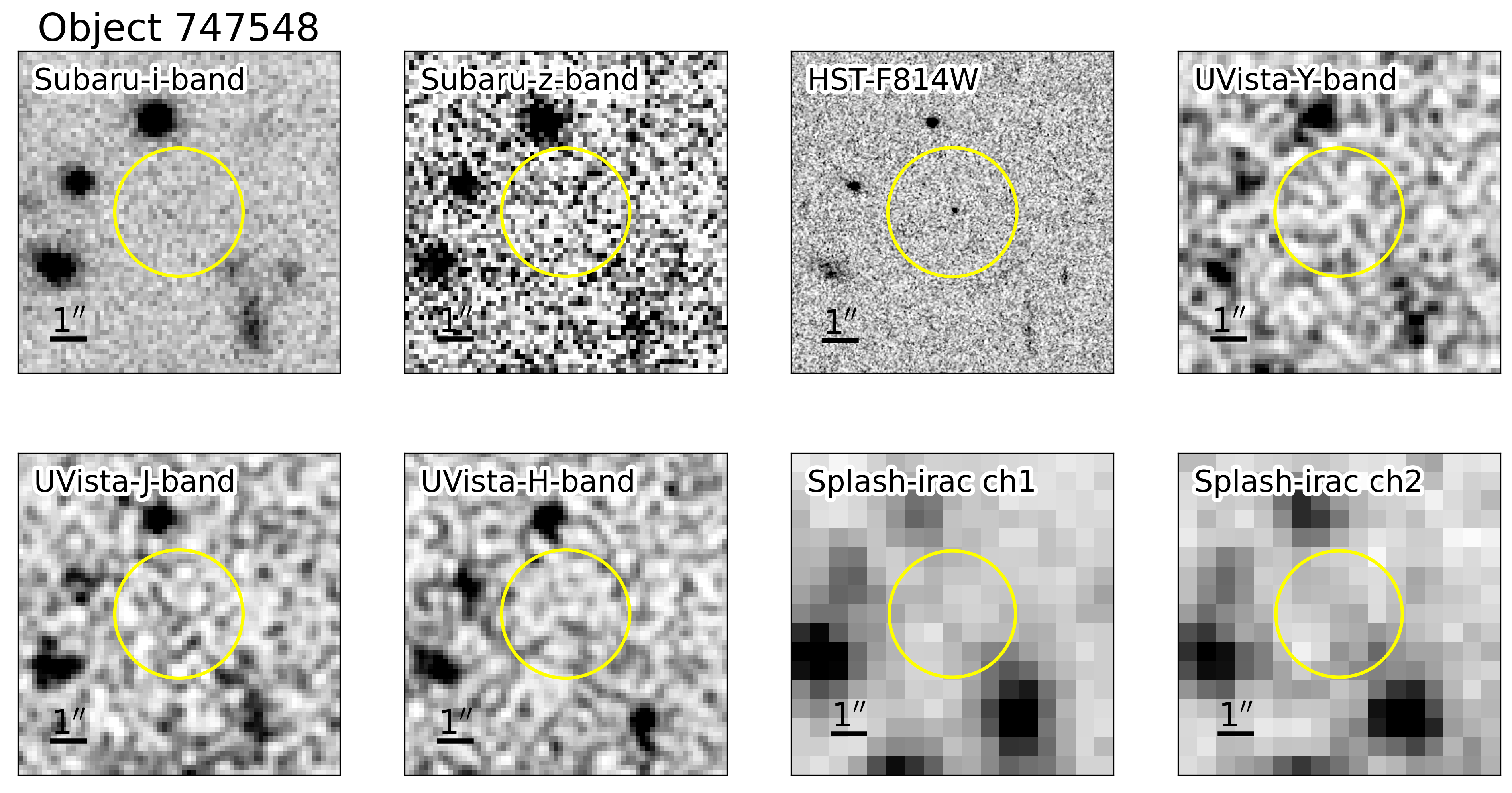}
\caption{  Images cutout of our sample.  For object 303826, we show the image cutout of F160W instead of H-band. 
\label{fig:other_source_img}}
\end{figure}

\section{Catalog of Grism Spectra}\label{sec:appendix1}

\begin{figure}[h!]
\centering
\includegraphics[width=0.80\linewidth]{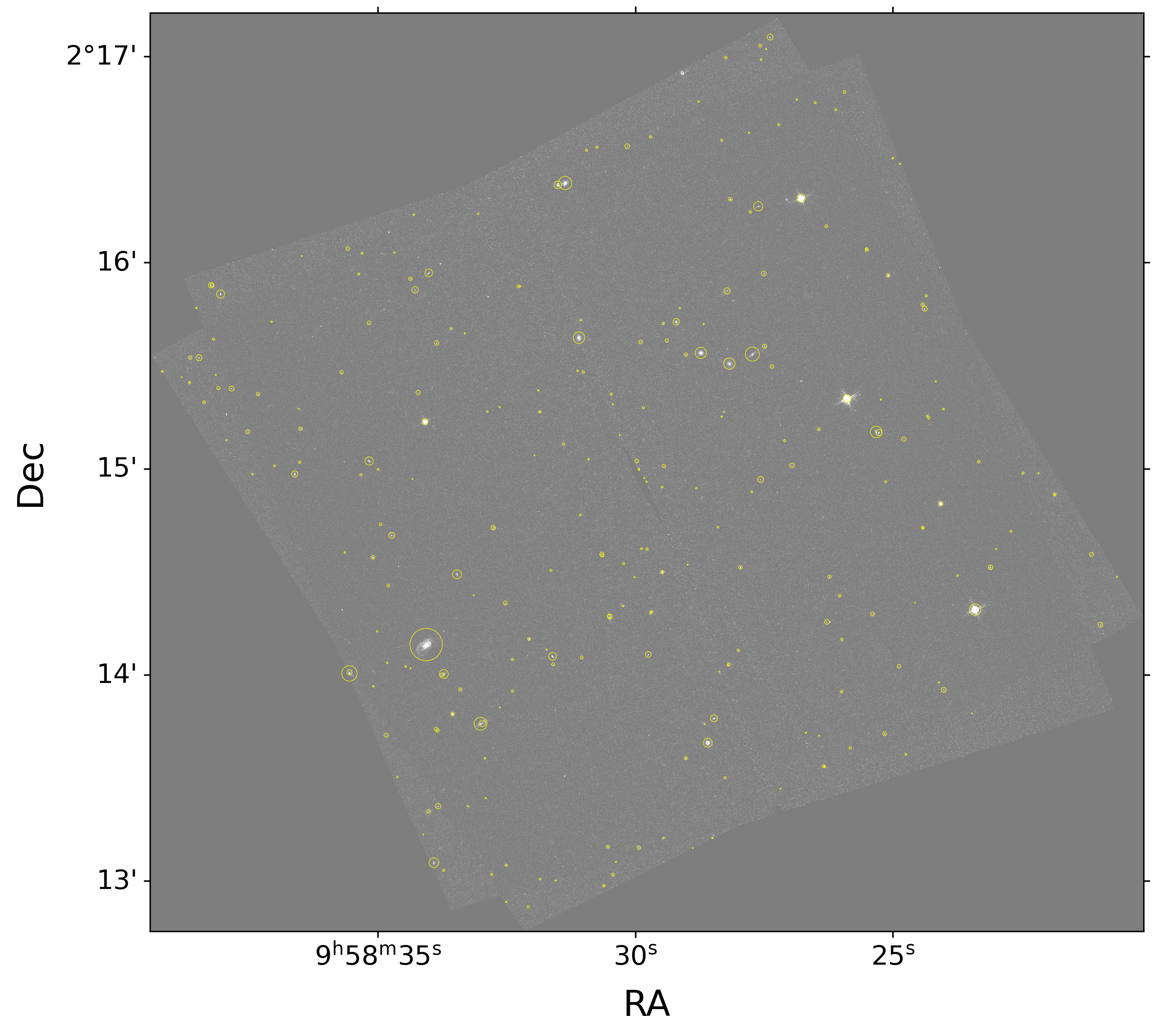}
\caption{  The direct image of the sky pointed to the target 667155. The sources with the spectra extracted are circled in yellow. 
\label{fig:direct_img}}
\end{figure}

We extract the grism spectra of the sources within the HST Cycle 30 GO proposal ID 17091.  We present the direct image in Figure~\ref{fig:direct_img} and the source catalog in Table~\ref{tab:grism_cat}, with the full table available online. Snapshots of the 2D grism and the extracted spectra are available on Zenodo: \dataset[10.5281/zenodo.15171451]{http://dx.doi.org/10.5281/zenodo.15171451}.  The spectra underlying this study are available from the corresponding author upon request.

\begin{table*}
\centering 
\begin{tabular}{lccccc} 
\hline\hline
name & RAdeg & DEdeg  & ID & photoz & flag   \\
\hline
667155$\_$00006 & 149.633672 & 2.214578 & 834165 & 0.9760 & 0  \\
667155$\_$00007 & 149.635447 & 2.214968	& 835620 & 1.0612 & 0  \\
667155$\_$00018 & 149.627529 & 2.216287	& 837521 & 0.6388 & 0  \\
667155$\_$00642 & 149.648139 & 2.233529	& 860496 & 0.5343 & 1  \\
\hline
\end{tabular}
\caption{ An subset of the source catalog as an example. The spectrum file names are labeled according to the AGN candidate target pointing they belong to. For instance, the sources shown here are part of the pointing for target 667155. We flag the spectrum if an emission line feature is presented.  The complete source table across all pointings is available online.  
}  
\label{tab:grism_cat} 
\end{table*}

%\section{Catalog of Grism Spectrum}\label{sec:appendix2}

\bibliography{sample631}{}
\bibliographystyle{aasjournal}

\end{document}